\documentclass[preprint,12pt,3p]{elsarticle}

\usepackage{amssymb}
\usepackage{graphicx}
\usepackage{mathrsfs}
\usepackage{amsmath}

\usepackage{color}

\newcommand{\ka}{\mathbf{k}}

\renewcommand\labelenumi{(\roman{enumi})}
\renewcommand\theenumi\labelenumi

\journal{Computational Materials Science}

\begin{document}

\begin{frontmatter}

\title{Bloch oscillations in two-dimensional crystals: Inverse problem}

\author[label1]{M Carrillo}
\address[label1]{Laboratorio de Inteligencia Artificial y Superc\'omputo, Instituto de F\'isica y Matem\'aticas, Universidad Michoacana de San Nicol\'as de Hidalgo, Morelia, 58040, M\'exico}
\ead{mcarrillo@ifm.umich.mx}

\author[label1]{J A Gonz\'alez}
\ead{gonzalez@ifm.umich.mx}

\author[label1,label2]{S Hern\'andez}
\address[label2]{Instituto de Ciencias Nucleares, Universidad Nacional Aut\'onoma de M\'exico, Apartado  Postal 70-543, Ciudad de M\'exico, 04510, M\'exico.}
\ead{sortiz@ifm.umich.mx}

\author[label1]{C E L\'opez}
\ead{clopez@ifm.umich.mx}

\author[label1]{A Raya}
\ead{raya@ifm.umich.mx}

\begin{abstract}
		Within an artificial neural network (ANN) approach, we classify {\em simulated signals} corresponding to the semi-classical description of Bloch oscillations on a two-dimensional square lattice. After the ANN is properly trained, we consider the inverse problem of Bloch oscillations (BO) in which a new signal is classified according to the lattice spacing and external electric field strength oriented along a particular direction of the lattice with an accuracy of 96\%. This approach can be improved depending on the time spent in training the network and the computational power available. This work is one of the first efforts for analyzing the BO with ANN in two-dimensional crystals.
\end{abstract}


\end{frontmatter}

\section{Introduction}
	
Flat two-dimensional crystals are unstable against thermal fluctuations according to the Mermin-Wigner theorem~\cite{Mermin}. Therefore, the early study of these crystals was considered just for academic convenience. More recently, it has been known, nevertheless, that some interesting phenomena occur effectively in two-dimensions, like quantum Hall effect~\cite{qhe1,qhe2} and high-$T_c$ superconductivity in cuprates~\cite{HTc}. Soon after the first isolation of graphene flakes~\cite{graphene1,graphene2}, a new era of materials science  emerged~\cite{graphene3} with a huge variety of two-dimensional (2D) systems discovered in the recent past~\cite{Weyl}. 
The 2D materials are nowadays a cornerstone of solid state physics and materials science because of their potential technological applicability and their impact in fundamental research. Many of these 2D crystals have the crystal structure of the square lattice, which due to its high symmetry, allows the study of a number of interesting phenomena, like Bloch oscillations (BO) \cite{Bloch}. It is well known that BO are not observed directly on crystals because of intraband tunneling and  ultrafast electron scattering; BO are directly observed in high purity superlattices under different experimental setups~\cite{exp1,exp2,exp21,exp22,exp3,exp31,exp4,exp41,exp5,exp51}. The equations of motion of BO are also relevant for a number of optical systems~\cite{Array1,Array2}. For that purpose, in a previous work~\cite{IBP}, some of us posed the inverse problem of BO for the linear chain within an artificial neural network (ANN) approach~\cite{ANN,dsp}. The idea is to use simulated signals for BO in a semiclassical approximation to train the ANN and then classify a new signal according to the lattice spacing and electric field strength with high accuracy. In this paper we extend these ideas to the 2D square lattice.
	
We develop a framework in which the ANN is trained using the simulated signals corresponding to the semiclassical description of BO for a 2D square lattice considering only the nearest neighbor influence. We then predict the strength of electric field along a particular direction of the lattice and the lattice spacing that produce such trajectories. We achieve up to $96\%$ of accuracy in our classification scheme, which can be improved depending on the computational time and computer power available. 
 
	For the presentation of ideas, we have organized the remaining of this paper as follows: In Section~\ref{sec:BO} we give a description of the BO phenomenology in the semiclassical approach. In Section~\ref{sec:signals}, we describe how the signals were generated and the ANN configuration. In Section~\ref{sec:results} the results for all the analyzed cases are discussed and finally, in Section~\ref{sec:conclusions}, the conclusions are presented.	
	
	\section{Bloch oscillation: Semiclassical approach}\label{sec:BO}
	We start our discussion from the tight-binding Hamiltonian of a monoatomic 2D square lattice of spacing $a$. Considering the nearest neighbors approximation, we have
	\begin{eqnarray}
		H \psi_{n,m}(\ka) &=& - t \psi_{n+1,m}(\ka) - t \psi_{n-1,m}(\ka)\nonumber\\
		 && - t \psi_{n,m+1}(\ka) - t \psi_{n,m-1}(\ka) + \epsilon_0 \psi_{n,m}(\ka)\nonumber\\
		&\equiv&\mathcal{E}^{(n,m)}(\ka) \psi_{n,m}(\ka)\;,
	\end{eqnarray}
	where $t$ is the hopping parameter and $\ka = k_1\hat{e}_x+k_2\hat{e}_y$ is the crystal-momentum of electrons in 2D. From Bloch theorem, it is straightforward to find that  the energy-momentum dispersion relation is:
	\begin{equation}
		\mathcal{E}^{(n,m)}(k_1,k_2) = \epsilon_0  - \epsilon^{(n,m)}(k_1,k_2)\;, 
	\end{equation}
	where
	\begin{equation}
		\epsilon^{(n,m)}(k_1,k_2) =  w (1 - \cos(k_1 a) - \cos(k_2 a)), 
		\label{en}
	\end{equation}
	$\epsilon_0$ is the on-site energy and $w=2t$. Next, we recall the semiclassical equations of motion  for an electron moving in an external electric field $\mathbf{E}$ oriented parallel to one direction of the square lattice, 
	\begin{eqnarray}
		\frac{d\mathbf{k}}{dt} & = & -e\mathbf{E}\;, \label{em1}\\
		\frac{d\mathbf{r}}{dt} & = & \frac{1}{\hbar}\frac{\partial}{\partial\mathbf{k}}\epsilon^{(n,m)}(k_1,k_2)\;.\label{em2}
	\end{eqnarray}
	%where instead of $\epsilon(k_1,k_2)$ we can use $\epsilon^{(n,m)}(k_1,k_2)$ given in Eq.~(\ref{en}).
	We can straightforwardly integrate the equations of motion and obtain the velocities and trajectories for a given external field strength. 
	Considering the lattice oriented along the $x-y$ plane and a uniform electric field $\mathbf{E}=E_1\hat{e}_x+E_2\hat{e}_y$, we integrate Eq.~(\ref{em1}) assuming the initial condition $k_j(0) = 0$ with $j = 1,2$. Thus
	\begin{equation}
		k_j(t)=-\frac{eE_j}{\hbar}t. 
	\end{equation}
	Rewriting Eq.~(\ref{em2}), the electron velocity is given by:
	\begin{eqnarray}
		v_{j}^{(n,m)}(k_j(t)) & =& \frac{w a}{\hbar}\sin(k_j(t) a),\nonumber\\
		&=& -\frac{w a}{\hbar}\sin\left(\frac{eE_ja}{\hbar}t\right),\label{veln}
	\end{eqnarray}
	and the electric current is simply $j_i=-ev_i$. Integrating Eqs. (\ref{veln}) we get the profile of BO obtaining the position of the electrons as function of time:
	\begin{eqnarray}
		x_j^{(n,m)}(t) & = & \frac{w}{eE_j}\cos\left(\frac{e E_j}{\hbar}a t\right),\nonumber \\
		& = & \frac{w}{eE_j}\cos(\omega_{E_j} t),
		\label{1v}
	\end{eqnarray}
	with $\omega_{E_{j}} =  e E_{j} a/ \hbar$. Eqs.~(\ref{1v}) describe the trajectories which are in fairly good agreement with the experimental observations of BO.  
	In the next Section we describe how the oscillations described by Eq. (\ref{veln}) are simulated and how ANN processes them in order to give an accurate result.

	\section{Signals creation and feature processing} \label{sec:signals}
	
	For fixed lattice parameters $a$ and $t$, the trajectories described by Eqs. (\ref{veln}) and (\ref{1v}) are functions of the electric field strength along each spatial direction, which becomes the only free parameter that characterizes a given trajectory in our considerations. We have trained an ANN that associates the electric currents of the electrons with their corresponding electric fields. In other words, the ANN learns through some examples the relationship between the electric current signals in the 2D square lattice and the electric fields that generate those currents. First, let us describe how the training signals were generated then we explain the classification process.
	
	For simplicity and without loss of generality, all signals were created following the next considerations:
	\begin{itemize}
		\item The parameters of Eqs. (\ref{veln}) and (\ref{1v}) were fixed to dimensionless units $e=\hbar=1$, $w_2 = w_2 = a = 0.5$.
		\item The signals were generated for a time lapse $\tau=200$.
		\item We integrate the signals considering the possibility of negative and positive electric fields for both $E_1$ and $E_2$ on three different ranges defined by $E_{\rm{min}}$ and $E_{\rm{max}}$. These cases will be describe more thoroughly later on section \ref{sub:cases}.
	\end{itemize}
	
\begin{figure}
		\centering
		\includegraphics[width=7cm]{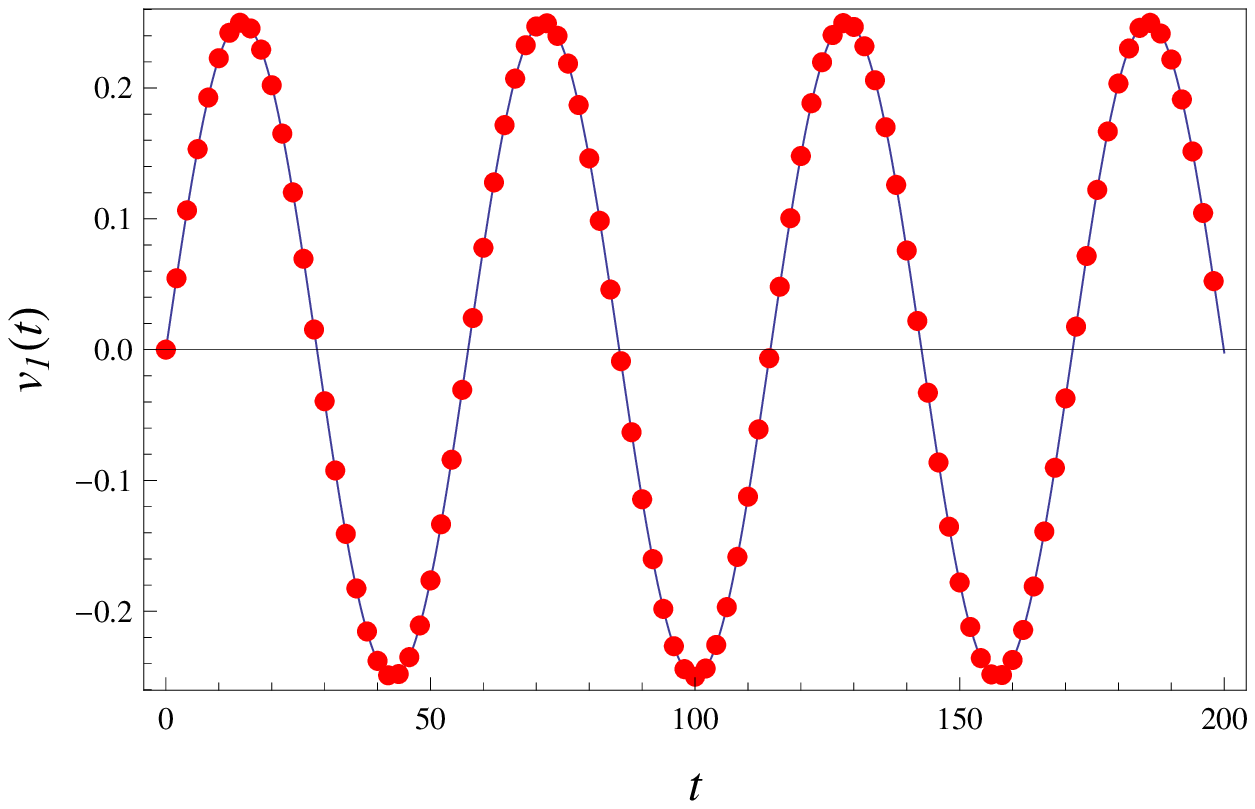}
		\includegraphics[width=7cm]{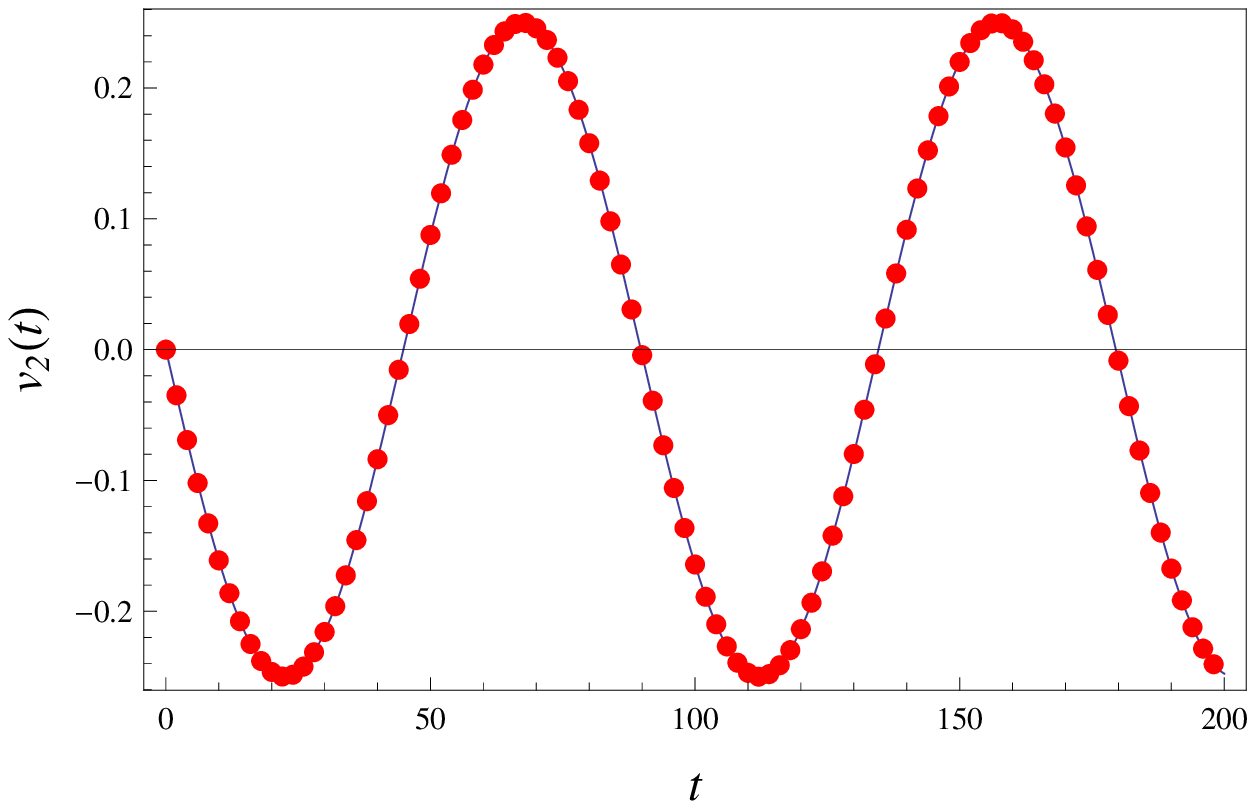}
		\caption{\label{vel} Velocities of the oscillating electrons generated using Eq.~(\ref{veln}). The points show the values used as inputs for the ANN. Left: velocity $v_1(t)$ for $E_1=-0.22$. Right: velocity $v_2(t)$ for $E_2=0.14$. }
	\end{figure}

	Once the signals were produced, we selected as inputs of the ANN values for each component of the velocity ($v_1$ and $v_2$) at one hundred different times defined by $t_i = i \Delta t$, with $\Delta t = \tau /100 = 2$ and $i = 0,1,\dots,99$. This means that the ANN will analyze a signal $V$ consisting of two hundred values:
	\begin{equation}
		V=\{v_1(t_1),v_2(t_1),\ldots,v_1(t_n),v_2(t_n) \}.
	\end{equation}
	\noindent
	In Figure \ref{vel} we show an example of BO velocities and the corresponding values where the trajectories were evaluated with $E_1 = -0.22$ and $E_2 = 0.14$ generated using Eq.~(\ref{veln}).
	
\begin{figure}
        \centering{\includegraphics[width=12cm]{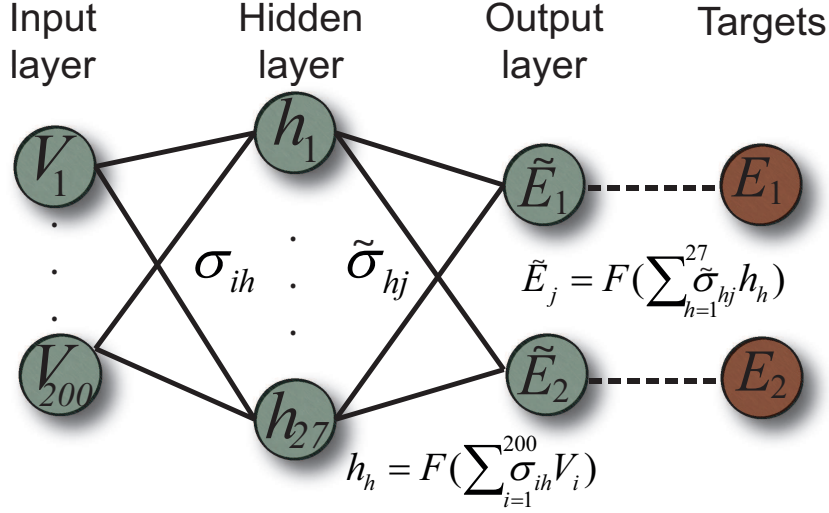}}
        \caption{Structure of the ANN developed for the classification of the BO. The input layer consists of 200 neurons according to the values extracted from the BO signals. The input layer is connected to a hidden layer with 27 units with weights $\sigma_{ih}$. Each hidden neuron computes a value using the sigmoid activation function $F$. Later, these values are sent to the 2 neurons in the output layer, with weights $\tilde{\sigma}_{hj}$ and using the same sigmoid function as before. Finally, the difference between from the ANN outputs and the proposed targets, associated to the electric field used in the BO signal is measured. With this difference the cost function is constructed and minimized.}
        \label{annstr}
    \end{figure}

	As the goal is to classify the electric field in 2D, we impose that the feedforward ANN has two outputs $\tilde{E}_1$ and $\tilde{E}_2$. Notice the difference between $\tilde{E_i}$ as the predicted value and $E_i$ the physical value. Considering a single hidden layer with 27 neurons, the equation that defines the predicted value given an input signal $V$ is defined by:
	\begin{equation}
		\tilde{E}_j = F\left(\sum_{h=1}^{27} \tilde{\sigma}_{hj}F\left(\sum_{i=1}^{200} \sigma_{ih} V_i \right) \right),
		\label{eq:ANNEQ}
	\end{equation}
	\noindent where $j = 1,2$. $F$ is the activation function for the hidden and output layers, in this case the standard sigmoid logistic function were used; $\sigma_{ih}$ and $\tilde{\sigma}_{hj}$ are the weights between the input and hidden layer and hidden to output layer respectively. The ANN structure is illustrated in the Figure \ref{annstr}.

	\subsection{Electric field scenarios}\label{sub:cases}
	
	The accuracy of the ANN depends on the frequency of the signals, the electric fields and sampling points. In this Section, we analyze how the performance of the ANN behaves in three different scenarios. Using 625 signals with all the parameters kept fixed except for the electric field that ranges in the scenarios:
	
	\begin{enumerate}
		\item Between  $[E_{\rm{min}} = -0.5, E_{\rm{max}} = 0.46]$ separated in steps of $\Delta E = 0.04$. \label{item:i}
		\item Between $[E_{\rm{min}} = -1, E_{\rm{max}}=0.92]$ with $\Delta E_j = 0.08$. \label{item:ii}
		\item Between $[E_{\rm{min}} = -0.25  , E_{\rm{max}} = 0.23]$ with $\Delta E_j = 0.02$.\label{item:iii}
	\end{enumerate}
	
\begin{figure}
	\includegraphics[width = \textwidth]{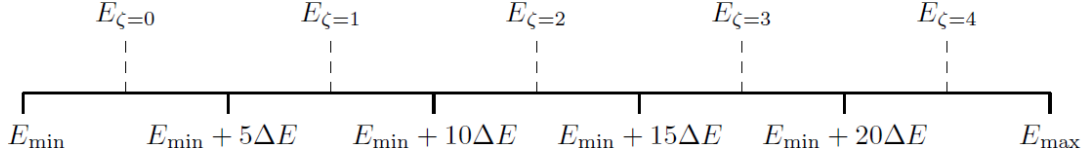}
	\caption{Schematic representation of the class definition for the electric field ranges. Also, the center of each region $E_\zeta$ is equivalent to the target class ($\hat{E}_\zeta$) that will be used in the training phase. The representation is shown only in one direction of the electric field but it is done in the same way in the other direction.}
	\label{fig:Schematics}
\end{figure}

	Considering that the activation function $F$ used in Eq. (\ref{eq:ANNEQ}) is a sigmoid function, the output of the network will be within the range $[0,1]$. The ANN's outputs could be divided in classes that represent the target intervals for $E_1$ and $E_2$. This means that the more classes an output has, the more precision is required for a correct classification. For this case, we have decided to divide each output in 5 classes. For clarity, let us develop the case (\ref{item:i}) where $\Delta{E_j} = 1/25$ and $E_{\rm{min}} = -0.5$ and $E_{\rm{max}} = 0.46$. Therefore for each $E_j$, every class covers up the range:
	\begin{equation}
		E_{\rm{min}} + 5 \zeta \Delta E \le E_\zeta < E_{\rm{min}} +  5 (\zeta + 1) \Delta E, \hspace{1cm}0\le \zeta \le 4,
	\end{equation}
	where $E_\zeta$ index each class for any of signal $E_j$ sections. An schematic representation classes division is presented in Figure \ref{fig:Schematics}. However, because the ANN's output is defined between (0,1), we need to map the electric field class classification into this range. For that, we define the center each one of the five classes $\hat{E}_\zeta$ in the output neuron as: 
	 \begin{equation}
		E_\zeta \equiv \hat{E}_{\zeta} =  0.1 + 0.2  \zeta.
		\label{clases}
	\end{equation}
	
	Besides, the center of each class will be used as the target value ($\hat{E}_\zeta$) in the training phase.	For example, if the signal is created with any of the first five values for $E_1$ ($\zeta = 0$) and the last five values of $E_2$ ($\zeta = 4$), then the ANN has correctly classified this signal if:
	\begin{eqnarray}
		\hat{E}_0 - 0.1 \le \tilde{E}_{1} < \hat{E}_0 + 0.1, \\
		\hat{E}_4 - 0.1 \le \tilde{E}_{2} < \hat{E}_4 + 0.1. 
	\end{eqnarray}
	In the following section we discuss the training procedure used to minimize the error of the predictions.
	
	\subsection{ANN's training considerations}
	Given that the ANN's weights are adjusted under a supervised training, thus, for each signal is necessary to associate the outputs with their respective electric field used during the signal generation. Using all previous considerations, the ANN was trained with an offline backpropagation algorithm by minimizing the cost function:
	\begin{equation}
		C(\vec{\sigma})  = \frac{1}{2 P}\sum_{p=1}^P c^p = \frac{1}{2 P}\sum_{p=1}^P \sum_{j=1}^2 {\left( \tilde{E}_{j}^{p} - \hat{E}_{j}^{p}\right)}^2.
	\end{equation}

	\noindent where $p=1,2,...,P$, with $P$ the number of training patterns. This backpropagation algorithm is a gradient descent method that adjusts the weights $\vec{\sigma}$ after an epoch or iteration $s$ by following the relationship: 
	\begin{equation}
		\sigma_{lm}(s+1)=\sigma_{lm}(s) - \frac{\gamma}{P}\sum_{p=1}^P\delta_{lm}^p(s),
	\end{equation}
	where $\gamma$ is the learning rate and the indexes $l \textrm{ and } m$ indicate the connection between the $l$ and the $m$ neuron and $\delta_{lm}^p$ is defined by

	\begin{equation}
		\delta_{lm}^p(s)= \frac{\partial{C(s,\vec{\sigma})}}{\partial{\sigma_{lm}^p(s)}}.
	\end{equation}
	
	The number of steps $S$ ($1 \le s \le S$) can be selected by achieving a default error, a maximum number of iterations or by cross-validation with an unknown set of signals. In this case, from the 625 signals generated, we chose randomly seventy percent of them as the training set ($P=438$) and the remaining 187 signals were used as the validation set to check convergence and avoid overfitting to the proposed targets. Moreover, another set of 187 signals were generated to test the accuracy of the ANN to completely unknown signals, namely the test set. During the training phase a maximum of ten thousand iterations were considered, all ANN's weights were initialized randomly between [-1,1] and a learning rate $\gamma=0.005$ was used. Moreover, to help the network to converge faster the cost function to a minimum, it is convenient that all the inputs have the same order of magnitude. Therefore a min-max normalization in the velocities is performed in every input:
	\begin{equation}
		\tilde{V}^p_i= \left\{
		    \begin{array}{rl}
			\frac{V^p_i-<V_i>}{V_{i}^{\rm{max}}-V_{i}^{\rm{min}}}, & \text{if}\hspace{.5cm} V_{i}^{\rm{max}}\neq V_{i}^{\rm{min}}\\
			V^p_i-<V_i>,& \text{if }\hspace{.5cm}V_{i}^{\rm{max}}=V_{i}^{\rm{min}}
    		\end{array}
    		\right.	\end{equation}
	for $1\le i \le 200$, $<V_i>$ is the average, $V_{i}^{\rm{max}}$ and $V_{i}^{\rm{min}}$ are the maximum and minimum values respectively of the $i-$th input of all the signals.

	\section{Results}\label{sec:results}
	
	In all the results presented below, we have trained five different networks using different initial weights, and reporting only the best one for all the cases described in the past section. 
	For the first case (\ref{item:i})  this ANN has classified correctly with a perfect score the training set, meanwhile with the test set it achieved a 82\% and 91\% efficiency on $\tilde{E}_1$ and $\tilde{E}_2$ respectively as it is observed in Table \ref{tab:Table}. In this case, the extreme ranges for $E$ are wider and, according to Eq.~(\ref{1v}), the maximum frequency of the signals is twice as in the case (\ref{item:i}). Therefore, the signals have a higher frequency, so in principle we should need more points or a lower interval in time to characterize properly these signals before being introduced into the ANN. We consider that the lower accuracy in the network is due to this fact.\\

	\begin{table}
	    \begin{center}
			\begin{tabular}{|c|c|c|c|c|c|c|}
				\hline 
				& \multicolumn{2}{c|}{Training set (\%)} & \multicolumn{2}{c|}{Validation set (\%)} & \multicolumn{2}{c|}{Test set (\%)}\tabularnewline
				\hline 
				\hline 
				Output & $\tilde{E}_{1}$ & $\tilde{E}_{2}$ & $\tilde{E}_{1}$ & $\tilde{E}_{2}$ & $\tilde{E}_{1}$ & $\tilde{E}_{2}$\tabularnewline
				\hline 
				Case (i) & 100 & 99 & 90 & 86 & 82 & 91\tabularnewline
				\hline 
				Case (ii) & 99 & 99 & 90 & 85 & 42 & 40\tabularnewline
				\hline 
				Case (iii) & 100 & 100 & 98 & 97 & 96 & 93\tabularnewline
				\hline 
			\end{tabular}
			
			\caption{Percent of correctly classified patterns of the ANN for the signals
				created varying: Case (i); $E_{1}$ between $\left[-0.5,0.46\right]$
				and $E_{2}$ between $\left[-0.5,0.46\right]$, Case (ii); $E_{1}$
				between $\left[-1,0.92\right]$ and $E_{2}$ between $\left[-1,0.92\right]$,
				Case (iii); $E_{1}$ between $\left[-0.25,0.23\right]$ and $E_{2}$
				between $\left[-0.25,0.23\right]$. We used 438 training patterns,
				187 validation patterns and the average of 5 test sets of 187 patterns
				each one.}
			\label{tab:Table}
		\end{center}
	\end{table}

	The results for case number (\ref{item:ii}) where the extreme values of $E_j$ were in the range [$-1,0.92$], are less accurate because the outputs only achieved an accuracy of $\tilde{E}_1 = 42\%$ and $\tilde{E}_2 = 40\%$ in the test set as shown in Table \ref{tab:Table}.
	
\begin{figure}
		\centering
		\includegraphics[width=7cm]{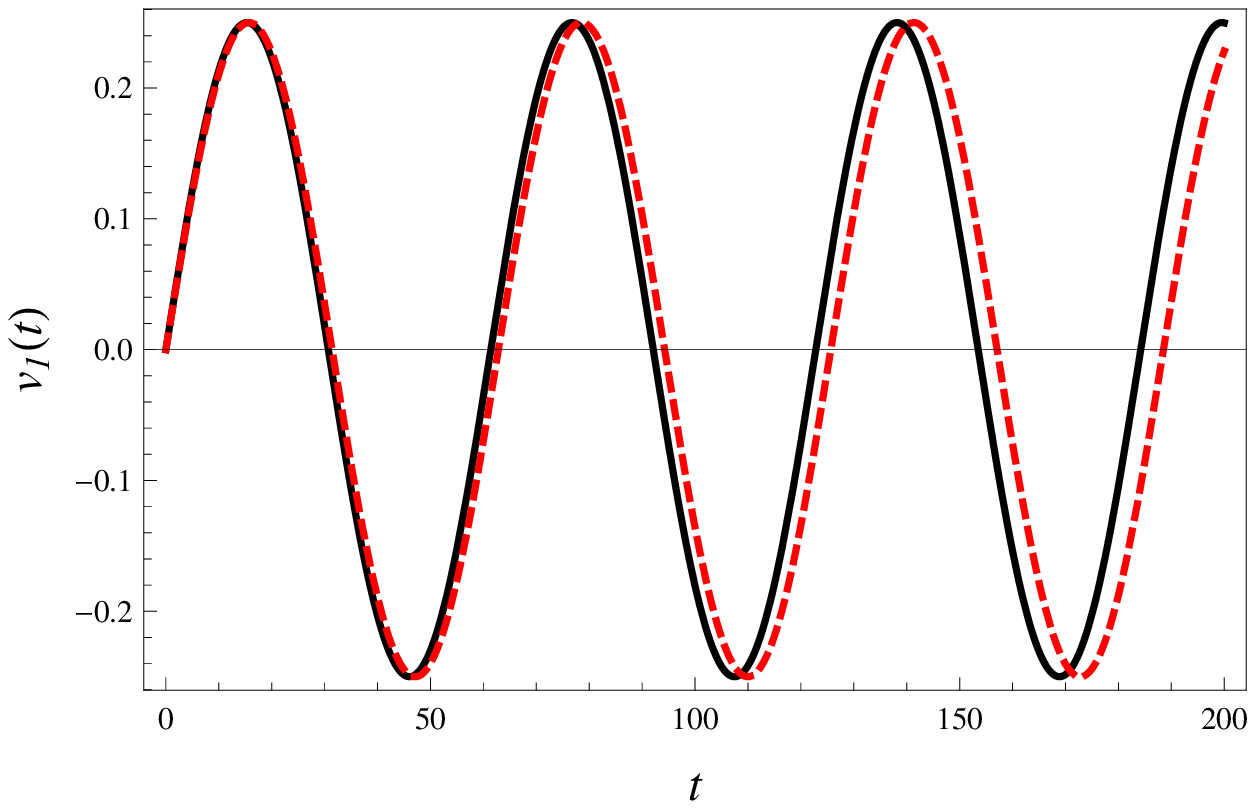}
		\includegraphics[width=7cm]{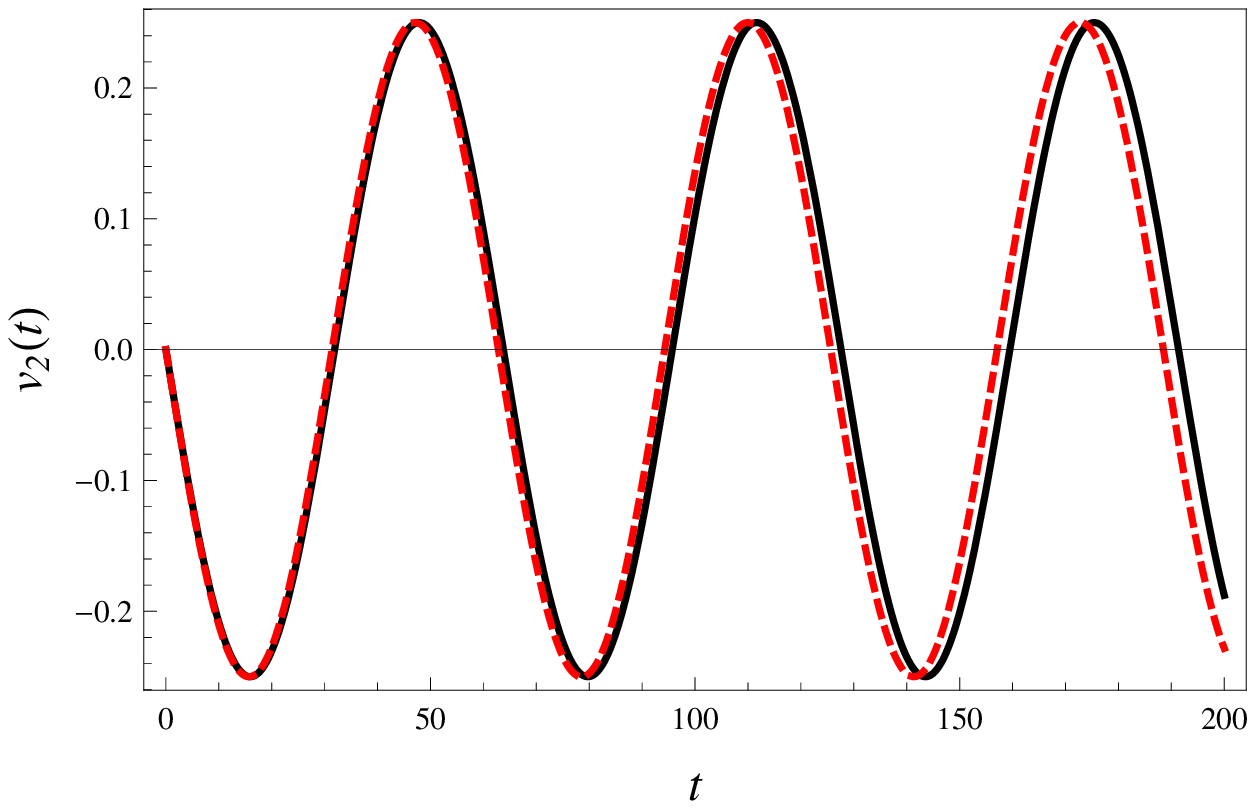}
		\caption{\label{velpre} Plot of $v_1(t)$ and $v_2(t)$, for the oscillations of the electrons generated with random values $E_1$ and $E_2$ between [-0.25,0.23]. The solid lines represent the BO oscillation generated using the true electric field $E_1$ and $E_2$, while the dash curves are the BO signals generated using the center values of the electric fields on the predicted classes $\tilde{E_1}$ and $\tilde{E_2}$, as described on Section~\ref{sub:cases}. On the left are plotted $v_1(t)$ for the true electric field $E_1=-0.2046$ and the center value of class 0, i.e, $E_{\zeta = 0} \equiv E_1= -0.20$. On the right are plotted $v_2(t)$ for $E_2 = 0.1969$ and the center value of $E_{\zeta = 4} \equiv E_2 = 0.2$. }
	\end{figure}

	Finally, for case (\ref{item:iii}) with the interval $[E_{\rm{min}}=-0.25, E_{\rm{max}}=0.23]$ and $\Delta E=1/50$, the curves generated have a lower frequency than the case (\ref{item:i}), thus the sampled points have more information about the signal, letting the ANN to outperform previous cases, were achieved an efficiency of $96\%$ and $93\%$ for $E_1$ and $E_2$ respectively, as it is shown in Table \ref{tab:Table}. An example of the prediction using the BO signals with an electric field composed by $E_1 = -0.2046$ and $E_2 = 0.1969$ is shown in Figure \ref{velpre}. In this case, the ANN estimates, that the signals where generated with values of $E_1$ and $E_2$ belonging to the classes $E_{\zeta = 0}$ and $E_{\zeta =4}$ respectively, which is a correct classification.
	
	\section{Final remarks}\label{sec:conclusions}
	We have developed a method employing an ANN approach to analyze the Bloch oscillations on a 2D square lattice of atoms within a tight-binding approximation considering the nearest neighbors influence. The ANN considered uses the velocity (electric current) oscillations signals as inputs signals and estimates the corresponding electric field strength projection along each spatial direction. For the purpose of this work, three different scenarios where the maximum and minimum electric fields considered are restricted. The extreme ranges of the electric fields determine the electron velocity frequencies and thus the number of points sampled per cycle, which impact the ANNs performance. The ANNs were trained and cross-validated with 625 signals within these ranges, meanwhile they were tested for signals with random electric fields on those same intervals. In the best case scenario, for low frequency, the ANN reaches at least 93\% accuracy on each output on the test set. As mentioned before, this is because for the lower interval of the electric field, the generated curves oscillate less and therefore the curves are described better. Meanwhile, for the greater interval of the electric field the predictions are less accurate, because the curves require more points to described them.

	From our previous work \cite{IBP} and the results exposed here, it is straightforward to see that this approach has a good potential and encourage us to explore more complex systems.
	
	\section*{Acknowledgments}
	We acknowledge support from CONACyT grant 256494 and CIC-UMSNH (MÂ´exico)
under grants 4.22 and 4.23.

%% References


\section*{References}
\bibliographystyle{elsarticle-num}

\end{document}